\documentclass[aps,prl,twocolumn,showpacs,floatfix]{revtex4}

\usepackage{dcolumn}
\usepackage{amsmath}
\usepackage{graphicx}

\begin{document}

\newcommand \be {\begin{equation}}
\newcommand \ee {\end{equation}}
\newcommand \bea {\begin{eqnarray}}
\newcommand \eea {\end{eqnarray}}
\newcommand \nn {\nonumber}
\newcommand \la {\langle}
\newcommand \rl {\rangle_L}

\title{Global fluctuations and Gumbel statistics}
\author{Eric Bertin}
\affiliation{Department of Theoretical Physics, University of Geneva,
CH-1211 Geneva 4, Switzerland}

\date{\today}

\begin{abstract}
We explain how the statistics of global observables in correlated systems
can be related to extreme value problems and to Gumbel statistics.
This relationship then naturally leads to the emergence of the generalized
Gumbel distribution $G_a(x)$, with a real index $a$,
in the study of global fluctuations. To illustrate these findings,
we introduce an exactly solvable nonequilibrium model describing
an energy flux on a lattice, with local dissipation,
in which the fluctuations of the global energy are precisely
described by the generalized Gumbel distribution.
\end{abstract}


\maketitle

The ubiquitous appearance of asymmetric distributions in the
study of fluctuations of global quantities
in correlated systems has raised a lot of interest in recent years.
Such non-Gaussian distributions, characterized by an exponential tail
on one side and a rapid fall-off on the other side, have been
observed in many models or experimental systems, in the context of turbulence
\cite{BHP,Pinton99,Fauve,Pinton02,Portelli03}, equilibrium critical systems
\cite{BHP,Bramwell-PRL,Bramwell-PRE,Portelli01,Clusel04a},
nonequilibrium models exhibiting self-organized criticality
\cite{Bramwell-PRL,Jensen}, interface models \cite{Racz94},
$1/f$ noise \cite{Antal}, Langevin equations \cite{Farago},
granular gas models \cite{Fauve,Brey},
or even the statistics of the level of the Danube river \cite{Danube}.
Quite strikingly, this analogy is not only qualitative, but many of the
distributions observed in these very different systems actually fall
\cite{BHP,Bramwell-PRL,Clusel04a,Brey,Danube},
once suitably rescaled, close to the so-called Bramwell-Holdsworth-Pinton
(BHP) distribution describing the magnetization of the XY model in the low
temperature limit, as well as the roughness of the two-dimensional
Edwards-Wilkinson surface model \cite{Bramwell-PRE,Goldenfeld}.
Yet, as not all data collapse onto the BHP curve \cite{Portelli01,Pinton02},
a more general distribution has been proposed to describe the data,
namely the generalized Gumbel distribution $G_a(x)$, which includes
a continuous shape parameter $a$
\cite{Bramwell-PRL,Bramwell-PRE,Portelli01,Portelli02,Danube,Pinton02,
Brey,Chapman}--see below for definition.
Interestingly, this distribution, with $a=1$, turns out to be the exact one
for periodic Gaussian $1/f$-noise \cite{Antal}; it is also very close to
the BHP one for $a \approx \pi/2$ \cite{Bramwell-PRL,Brey}.

The distribution $G_a(x)$ originates, for integer values of $a$,
from the study of extreme value statistics \cite{Gumbel,BouchMez},
and describes the fluctuations of the $a^{\rm th}$ largest value
in a large set
of identically distributed (independent) random variables $z_i$
\footnote{This requires that $z$ has no upper bound, and
that $P(z)$ decays faster than any power law at large $z$.}.
Accordingly, there is no obvious theoretical motivation for the use of
the distribution $G_a(x)$ in the study of fluctuations of global quantities.
Rather, it is usually considered as a convenient fitting function,
and a theoretical understanding of its relevance is still lacking.
Indeed, the question of the underlying role of extreme values in
correlated systems has been repeatedly asked in the literature
\cite{Antal,Jensen,Portelli02,Clusel04a,Racz03}.
Still, attempts to identify an extremal process dominating the dynamics
of such systems have failed up to now \cite{Portelli02,Clusel04a}.
All the above body of results thus leads to the following questions.
First, what is (if any) the precise relationship between global
fluctuations in correlated systems and extreme value statistics?
Second, could one find a simple physical model for which the
fluctuations of a global quantity would be exactly described by a
generalized Gumbel distribution?

Global fluctuations in complex correlated systems are often hard to tackle
analytically precisely due to strong correlations
between local microscopic variables. Yet, in some cases,
statistically independent collective variables --like Fourier modes
\cite{Villain,Bramwell-PRE,Clusel-TBP,Antal}-- can be defined,
so that a problem of correlated random variables may be converted into a
problem of independent random variables, with non-identical distributions
--otherwise the central limit theorem would hold.

In this Letter, we explain how the statistics of global quantities,
expressed as sums of non-identically distributed random variables,
is related to extreme value problems, and how the generalized
Gumbel distribution $G_a(x)$, with a real index $a$, emerges in
the study of global fluctuations.
Interestingly, it turns out that such a relationship does not rely on
an extremal process hidden in the dynamics of global variables,
contrary to usual conjectures.
These results are illustrated on a nonequilibrium cascade model
in which the fluctuations of the total energy are exactly
described by the generalized Gumbel distribution $G_a(x)$, where $a$
depends continuously on the microscopic parameters of the model.

Our starting point is the observation \cite{Antal}
that the integrated power spectrum $w$ of periodic Gaussian $1/f$ noise
is distributed, after a suitable rescaling $x=(w-\langle w \rangle)/\sigma_w$
(where $\sigma_w^2$ is the variance of $w$)
according to the Gumbel distribution $G_1(x)$.
The model for $1/f$ noise used in
\cite{Antal} consists in a large number $N$ of statistically independent
Gaussian Fourier modes with complex amplitudes $c_n=c_{-n}^*$.
Introducing $y_n \equiv |c_n|^2+|c_{-n}|^2$, one has by definition
$w = \sum_{n=1}^N y_n$. The distribution of $y_n$ reads
\be \label{pn_yn}
p_n(y_n) = n\kappa\, e^{-n\kappa\, y_n},
\ee
so that $w$ is simply
the sum of $N$ independent random variables $y_n$,
with non-identical exponential distributions.
Note that the appearance of a non-Gaussian distribution is not surprising
in itself, since the sum of the variances of the $y_n$'s converges
when $N \to \infty$ [as ${\rm var}(y_n)=1/(n^2\kappa^2)$], so that the central
limit theorem is not expected to hold --for a detailed discussion on this
point, see \cite{Feller}.
Still, the fact that a Gumbel distribution precisely emerges may
suggest the existence of some ``hidden'' extremal processes
dominating the fluctuations of $w$, but no clear evidence for such processes
has been found yet \cite{Portelli02,Clusel04a}.

\begin{figure}[t]
\centering\includegraphics[width=6.5cm,clip]{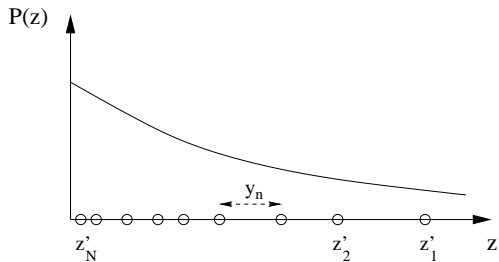}
\caption{\sl Sketch of the notations used in the text. $N$ random values of
$z$ are drawn according to the probability density $P(z)$. These values are
relabelled into $z_1'\ge \ldots\ge z_N'$, and the interval between two
successive $z_n'$ is denoted by $y_n$.
}
\label{fig-EVS}
\end{figure}

Actually, a different perspective may be necessary to understand
the relationship between the two problems. Indeed, instead of
looking for extremal processes hidden in sums of random variables,
one may look for sums of random variables with decreasing
amplitudes when studying the statistics of extreme values.
To this aim, we introduce the following procedure,
illustrated in Fig.~\ref{fig-EVS}.
Considering a set of $N$ random variables $z_n>0$ ($1\le n\le N$),
all drawn from the same distribution $P(z)$, we introduce
the variables $z_n'$ defined by ordering the original variables $z_n$:
$z_n'=z_{\sigma(n)}$, where $\sigma(n)$ is a permutation over the interval
$[1,N]$ such that $z_1'\ge z_2'\ge \ldots \ge z_N'$. Thus $z_1'$ is simply
the maximum value of the set $\{z_n\}$.
We also define the interval $y_n$ between $z_{n+1}'$ and $z_n'$:
\be
y_n = z_n' - z_{n+1}' \quad (1 \le n \le N-1) ; \quad y_N = z_N'
\ee
With these notations, one can write:
\be \label{max-sum}
\max_{1\le n\le N} (z_n) \equiv z_1' = \sum_{n=1}^N y_n
\ee
As a result, a problem of extreme value can be mapped onto a problem of sum of
random variables. Still, it should be noticed that
although the original variables $z_n$ are independent, the
$y_n$'s are a priori correlated.

In the following, we show that in the specific case where $P(z)$ is an
exponential distribution $P(z)=\kappa\, e^{-\kappa z}$,
the $y_n$'s prove independent and distributed according
to Eq.~(\ref{pn_yn}). The distribution $P_N(\{y_n\})$ reads:
\bea \nonumber
P_N(\{y_n\}) &=& \kappa^N N! \int_0^{\infty} dz_N \,e^{-\kappa z_N} \dots
\int_{z_2}^{\infty} dz_1 \, e^{-\kappa z_1} \\
&\times& \delta(y_N-z_N) \prod_{n=1}^{N-1}
\delta\left[y_n-(z_n-z_{n+1})\right]
\label{p-yn-integ}
\eea
where the integral over $z_n$ is from $z_{n+1}$ to $\infty$,
for $1\le n \le N-1$.
This expression can be understood as follows:
either the variables $\{z_n\}$ are
already ordered, which straightforwardly gives the above integrals,
or they are not, and then can be ordered through a permutation,
which leads to the $N!$ factor in front.
Making the change of variables $v_n=z_n-z_{n+1}$ ($1\le n\le N-1$)
and $v_N=z_N$ in Eq.~(\ref{p-yn-integ}),
the different integrals factorize and one finds:
\be \label{p-yn-prod}
P_N(\{y_n\}) = \prod_{n=1}^N n\kappa\, e^{-n\kappa y_n}
\ee
Thus it turns out that in the specific case $P(z) = \kappa e^{-\kappa z}$
the $y_n$'s are independent variables, distributed as
the squared Fourier amplitudes in the $1/f$ noise model, i.e., according to
Eq.~(\ref{pn_yn}).
But as the sum of the $y_n$'s is precisely
the maximum value of a set of exponentially
distributed variables $\{z_n\}$, we know that this sum has to be distributed
(after a suitable rescaling)
according to $G_1(x)$, so that one recovers immediately the results of
\cite{Antal}. Accordingly, a clear relationship appears
between the statistics of extreme values and that
of sums of variables with decreasing amplitudes.
This relationship can actually be understood at two different levels.
On the one hand, starting from a set of (possibly correlated) variables
$\{z_n\}$, one can always define the interval $y_n$
between two successive variables $z_n'$ obtained
by ordering the set $\{z_n\}$ --see Eq.~(\ref{max-sum}).
Thus, on general grounds, the maximum value of correlated variables
$z_n$ can be formally written as a sum of correlated variables $y_n$,
but the corresponding extreme value distribution is usually unknown.
On the other hand, it seems that the maximum value of a set $\{z_n\}$ of
{\it independent} variables is related to a sum of {\it independent} variables
$\{y_n\}$ only in the case where the $z_n$'s are drawn from an exponential
distribution, leading to the Gumbel distribution $G_1(x)$.
Indeed, the factorization property of the exponential is
essential to derive Eq.~(\ref{p-yn-prod}) from Eq.~(\ref{p-yn-integ}).

The above result leads to some rather unexpected consequences.
From the very definition of the variables
$\{z_n'\}$, $z_k'$ is precisely the $k^{\rm th}$
largest value of the original set $\{z_n\}$.
So we know that $z_k'$ follows, once rescaled as
$x=(z_k'-\langle z_k'\rangle)/\sigma_k$ with $\sigma_k^2={\rm var}(z_k')$,
the generalized Gumbel distribution $G_k(x)$ \cite{Gumbel}.
The distribution $G_a(x)$ is defined for any positive real value $a$ by:
\be
G_a(x) = \frac{\theta_a a^a}{\Gamma(a)} \exp\left\{-a\left[\theta_a(x+\nu_a)
+e^{-\theta_a(x+\nu_a)}\right]\right\}
\ee
with
\be
\theta_a^2 = \frac{d^2\ln\Gamma}{da^2}, \qquad
\nu_a = \frac{1}{\theta_a} \left(\ln a - \frac{d\ln\Gamma}{da}\right)
\ee
where $\Gamma(a)$ is the Euler Gamma function.
Besides, $z_k'$ may also be expressed as a sum:
\be
z_k' = \sum_{n=k}^N y_n = \sum_{n=1}^{N-k+1} \tilde{y}_n
\ee
with $\tilde{y}_n \equiv y_{n+k-1}$ distributed according to
\be \label{eq-dist}
p_{n,k}(\tilde{y}_n) = (n+k-1)\kappa\, e^{-(n+k-1)\kappa\tilde{y}_n}
\ee
Thus the sum of independent random variables drawn from
(\ref{eq-dist}) is distributed, after a suitable rescaling,
according to $G_k(x)$ in the limit $N \to \infty$.
But then, one can forget the original extreme value
problem, and consider only the statistics of the sum,
so that there is no more reason to restrict $k$ to be integer.
Since the generalized Gumbel distribution is
obtained for integer $k$, it seems plausible that it also holds
for real values $k=a>0$.
To be more specific, considering independent variables $u_n$ with 
distribution
\be \label{pn-alpha}
p_{n,a}(u_n) = (n+a-1)\kappa\, e^{-(n+a-1)\kappa u_n},
\quad 1 \le n \le N
\ee
the sum $X=\sum_{n=1}^N u_n$ is precisely distributed,
in the limit $N \to \infty$,
according to the generalized Gumbel distribution $G_a(x)$,
where $x=(X-\langle X \rangle)/\sigma_X$.
This result, suggested by the above argument, can be
derived exactly without reference to the extreme value problem \cite{long}.

\begin{figure}[t]
\centering\includegraphics[width=6cm,clip]{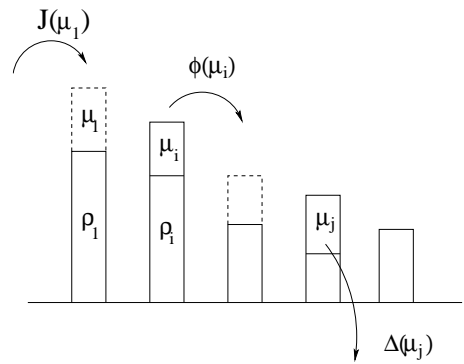}
\caption{\sl Schematic view of the model, with the three different mechanisms:
injection on the leftmost site with rate $J(\mu_1)$, transport
from site $i$ to site $i+1$ with rate $\phi(\mu_i)$, and dissipation on
site $j$ with rate $\Delta(\mu_j)$.
}
\label{fig-model}
\end{figure}

We now illustrate the above result on a simple nonequilibrium stochastic
model, which is defined by the following rules
\footnote{The present model is inspired by, but still quite different from,
cascade models for turbulence (see e.g.~\cite{Portelli03}).
It should rather be thought of as a generic model with boundary injection
and bulk dissipation.}.
On each site $n=1,\ldots, N$ of a one-dimensional lattice,
a positive continuous
variable $\rho_n$ --to be thought of as an energy-- is introduced. The
(asynchronous) dynamics is defined through three different
physical mechanisms involving energy,
namely injection on --say-- the left boundary, transport from one site
to its right neighbor, and local dissipation.
More precisely, an amount of energy between $\mu$ and $\mu+d\mu$
can be either injected on the leftmost site $n=1$
with a rate (probability per unit time) $J(\mu) d\mu$,
transferred from site $n$ to site $n+1$ with rate $\phi(\mu) d\mu$,
or removed (i.e., dissipated) from site $n$ with rate $\Delta(\mu) d\mu$
 --see Fig.~\ref{fig-model}.
On the rightmost site $n=N$, the transferred energy is actually dissipated.
Note that the above rates do not depend on the values of the local energies
$\rho_n$, apart from the obvious constraint
that one cannot withdraw from site $n$ (either for transport
or dissipation) an energy $\mu$ greater than $\rho_n$.

\begin{figure}[t]
\centering\includegraphics[width=7.5cm,clip]{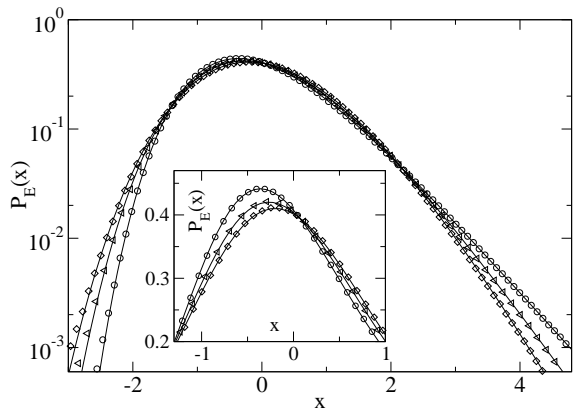}
\caption{\sl Distribution $P_E(x)$ of the rescaled (global) energy of
the model, $x=(E-\langle E \rangle)/\sigma_E$.
The analytical result $G_a(x)$ (full lines) is compared with
numerical simulations with $\phi(\mu)=1$, for $a \equiv 1+\beta/\lambda=1.7$
($\circ$), $3.3$ ($\triangleleft$), and $6$ ($\diamond$),
showing an excellent agreement. Inset: zoom on the top of the curves,
on a linear scale.
}
\label{fig-dist}
\end{figure}

The master equation governing the evolution of the probability distribution
$P(\{\rho_n\},t)$ reads:
\bea \nonumber
\frac{\partial P}{\partial t} &=&
\int_0^{\rho_1} d\mu J(\mu) P(\{\rho_1-\mu,\rho_j\},t)\\
\nonumber
&-& \int_0^\infty d\mu J(\mu) P(\{\rho_j\},t)\\
\nonumber
&+&\sum_{n=1}^{N-1} \int_0^{\rho_{n+1}} d\mu \phi(\mu)
P(\{\rho_n+\mu,\rho_{n+1}-\mu,\rho_j\},t)\\
\nonumber
&+&\sum_{n=1}^N \int_0^{\infty} d\mu [\Delta(\mu)+\phi(\mu) \delta_{n,N}]
P(\{\rho_n+\mu,\rho_j\},t)\\
&-& \sum_{n=1}^N \int_0^{\rho_n} d\mu [\phi(\mu)+\Delta(\mu)]
P(\{\rho_j\},t)
\eea
where $\rho_j$ generically stands for all the variables that are not
affected by $\mu$. In the following, we focus
on the specific case where $J(\mu)=e^{-\beta \mu} \phi(\mu)$
and $\Delta(\mu)=(e^{\lambda\mu}-1)\, \phi(\mu)$,
introducing two positive parameters $\beta$ and $\lambda$.
With these assumptions, the steady-state distribution $P(\{\rho_n\})$
proves factorized and can be computed exactly
\footnote{Technical details, as well as results for more general $J(\mu)$ and
$\Delta(\mu)$, will be reported elsewhere \cite{long}};
it turns out to be precisely the same as Eq.~(\ref{pn-alpha}):
\be
P(\{\rho_n\}) = \prod_{n=1}^N (\lambda n+\beta) \,
e^{-(\lambda n+\beta)\rho_n}
\ee
with the identification $\lambda=\kappa$ and $\beta=(a-1)\kappa$ --note that
$P(\{\rho_n\})$ does not depend on the specific form of $\phi(\mu)$.
As a result, the fluctuations of the total energy $E = \sum_{n=1}^N \rho_n$
are described in the infinite $N$ limit
--after rescaling $E$ to ensure zero mean and unit variance--
by the generalized Gumbel distribution $G_a(x)$, with $a=1+\beta/\lambda$
(see Fig.~\ref{fig-dist}).
Interestingly, in the limit of low dissipation $\lambda \to 0$,
one recovers a Gaussian distribution, since $G_a(x)$ converges
to a Gaussian for $a \to \infty$.
Qualitatively, the parameter $a$ may be thought of as the number of sites
having roughly the same energy, of the order of $1/\beta$.

Finally, we note that the `cascade' mechanism illustrated by the
above stochastic model should be considered as one possible mechanism,
but perhaps not as the unique one.
Indeed, in some systems like freely evolving granular gases \cite{Brey},
Gumbel distributions are indeed observed even though
the global quantity of interest cannot be written in an obvious way
as a sum of independent collective variables.
Yet, it must be noticed that such a granular system does not reach a
steady state since no energy is injected; fluctuations are then measured in
a scaling regime where the average kinetic energy continuously decreases.
Accordingly, one might expect another physical mechanism to
be at play in this case.

In summary, we have shown that the generalized Gumbel distribution $G_a(x)$
appearing in numerous experimental and numerical studies should not be
interpreted as a signature of some hidden extremal process, but on the
contrary, as the distribution associated to an infinite sum of independent
and exponentially distributed random variables $u_n$ ($n\ge 1$),
with mean value $[(n+a-1)\kappa]^{-1}$. If $a$ is integer, the variables $u_n$
can be interpreted as the intervals $y_n$ between two successive (ordered)
random values drawn from an exponential distribution
$P(z)=\kappa e^{-\kappa z}$, so that the $a^{\rm th}$ largest value
among the $z_n$'s
can be written as the sum of the $y_n$'s for $n \ge a$.
Thus a clear connection between global fluctuations
and extreme value statistics has been established.
Besides, we have proposed a simple nonequilibrium model, defined through
microscopic stochastic rules, for which the natural global quantity
is exactly described by the generalized Gumbel distribution $G_a(x)$,
with $a>1$ a real value related to the parameters of the model.
Such a simple model might be considered as a kind of `ideal' model,
that may be extended in several directions to describe in a more precise
way some realistic systems. In particular, one expects that changing
slightly the dynamical rules should yield a global energy distribution
which is still close to a Gumbel distribution.
In addition, the present model may be useful to study other issues
of nonequilibrium statistical physics, as there are very few known solvable
models including dissipation.

The author is grateful to I.~Bena, M.~Clusel, O.~Dauchot, M.~Droz,
P.~Holdsworth, C.~Mazza, F.~van Wijland and Z.~R\'acz for fruitful
discussions and interesting comments on the manuscript.
This work has been partially supported by the Swiss National Science
Foundation.

\end{document}